\documentclass[a4paper, 11pt, openany, oneside]{article}
\usepackage[cp1251]{inputenc} 
\usepackage[english]{babel} 
\usepackage{ graphics,amsfonts, amsmath,bm,amssymb, array, eucal}
\usepackage[dvips]{graphicx}
\usepackage[flushleft,small]{caption2}
\newcommand{\mc}[1]{\mathcal{#1}}
\newcommand{\E}{\mc{E}}
\makeatletter

\renewcommand{\@biblabel}[1]{#1.\hfill}
\newcommand{\diag}{\rm \diag\, }

\makeatother {

\begin{document}
\thispagestyle{empty}
\renewcommand{\abstractname}{\ }
\renewcommand{\refname}{\begin{center} REFERENCES\end{center}}
\newcommand{\const}{\mathop{\rm const\, }}

 \begin{center}\large
\bf LONGITUDINAL  DIELECTRIC PERMETTIVITY OF QUANTUM MAXWELL COLLISIONAL
PLASMAS
\end{center}
\begin{center}\large
  \bf  A. V. Latyshev$^1$ and A. A. Yushkanov$^2$
\end{center}

\begin{center}
{\it $^1$Department of Mathematical Analysis and Geometry,\\
{\rm \normalsize Electronic address: {avlatyshev@mail.ru}},\\
$^2$Department of Theoretical Physics,\\
{\rm \normalsize Electronic address: {yushkanov@inbox.ru}},\\
Moscow State Regional
University,  105005,\\ Moscow, Radio str., 10A}
\end{center}

 \begin{abstract}\large

The kinetic equation of Wigner -- Vlasov -- Boltzmann with
collision integral in relaxation BGK (Bhatnagar, Gross and Krook) form
in coordinate space  for quantum non--degenerate (Maxwellian)
collisional plasma is used.
Exact expression (within the limits of considered model) is
found.
The analysis of longitudinal dielectric permeability is done.
It is shown that in the limit when Planck's constant tends to
zero of expression for dielectric permettivity transforms
into the classical case of dielectric permettivity.
At small values of wave number it has been received
the solution of the dispersion equation.
Damping of plasma oscillations has been analized.
The analytical comparison with the dielectric Mermin' function
received with the use of the kinetic equation in momentum space is
done.
Graphic comparison of the real and imaginary parts of
dielectric permettivity of quantum and classical plasma is done also.

{\bf Key words:} collisional plasma, BGK equation,
electric conductivity, dielect\-ric permittivity, Lindhard's
formula, Landau's damping.

PACS numbers:  52.25.Dg Plasma kinetic equations,
52.25.-b Plasma proper\-ties, 05.30 Fk. Fermion systems and
electron gas
 \end{abstract}
 \Large
\begin{center}
\bf{1. Introduction}
\end{center}

In the present work formulas  for electric
conductivity and for dielectric permettivity of quantum
electronic non--degenerate Maxwellian plasma are deduced.

Dielectric permettivity in the collisionless quantum
gaseous plasma was studied by many authors (see, for example,
\cite {Klim}-\cite {Arnold}).
G. Manfredi  \cite{Manf} inves\-ti\-ga\-ted  one-dimensional
case of the quantum
plasma. In this article G. Manfredi has noted the importance of
carrying out of analysis of the dielectric permettivity deduced
with help of the quantum kinetic equation with collision integral
in  coordinate space.
The present work is devoted to performance of this  problem
for Maxwellian plasma.

In the present work for  derivation of dielectric permettivity
the quantum kinetic Wigner --- Vlasov --- Boltzmann equation
(WVB--equation) with  colli\-si\-on integral
in the form of $\tau$--models in coordinate space is applied.
Such collision integral is named the BGK collision integral.

The WVB--equation is written for Wigner function,
which is an analo\-gue of distribution function of
electrons for quantum plasma
(see \cite{Wigner}, \cite{Hillery} and \cite{Kozlov}).

The most widespread method of investigation of quantum plasma is the
method of Hartree --- Fock  or a method equivalent to it,
namely, the method of Random Phase Approximation \cite {Lif}, \cite {Plaz}.
In the work \cite {Lind} this method  has been applied for receiving
of the expression for
dielectric permettivity of quantum degenerate plasma in $ \tau $--approach.
However, in the work \cite {Kliewer} it is shown, that  expression
received in \cite {Lind} is noncorrect,
as it does not turn into classical expression under a condition, when
quantum amendments can be neglected. Thus in the work \cite {Kliewer}
empirically corrected expres\-sion  for
dielectric permettivity of quantum plasma, free from
the specified lack has been offered.
By means of this expression authors investigated
quantum amendments to optical proper\-ties of metal
\cite {Kliewer2}, \cite {Kliewer3}.
Friedel's oscillations in quantum plasma also have been inves\-ti\-gated
already for more than half a century (see, for example, \cite {Kohn1}-\cite {Harr}).

In the theory of quantum plasma two essentially
different possibili\-ties of construction of the relaxation kinetic
equation in $\tau$ - approach exist: in momentum space
(in the space of Fourier images of distribution function) and in
coordinate space.
On the basis of the relaxation kinetic
equation in the space of momen\-tum Mermin \cite {Mermin}
has carried out consistent derivation of the dielectric
permea\-bility for quantum collisional plasma in 1970 for the first time.

In the present work expression for the longitudinal
dielectric permettivity for non--degenerate plasma
with the use of the relaxation equations in space of coordinates is deduced.
If in the received expression we make
Planck constant converge to zero ($ \hbar\to 0$), we will receive
exactly the classical expression of dielectric permettivity of
non--degenerate plasma. Various limiting cases of the
dielectric permettivity are investigated.
Comparison with Mermin's result is carried out also.

\begin{center}
\bf{2. Solution of the kinetic equation}
\end{center}

We consider the kinetic Wigner --- Vlasov --- Boltzmann
equation \cite{Gurov} with collisional integral in the form of
BGK--model
$$
\frac{\partial f}{\partial t}+\textbf{v}\frac{\partial f}{\partial
\textbf{r}}=\dfrac{ie}{\hbar}W[f]+
\nu[f_{eq}(\mathbf{r},p,t)-f(\mathbf{r},\mathbf{p},t)].
\eqno{(2.1)}
$$
This equation describes evolution of the Wigner function
for electrons in quantum plasma.

Here $e$ is the charge of electron, $\hbar$ is the Planck's constant,
$\nu$ is the effective collision
frequency of electrons with ions and neutral atoms,
$f(\mathbf{r},\mathbf{p},t)$ is the Wigner function for electrons.

The function  $f_{eq}(\mathbf{r},p,t)$ is the equilibrium
Maxwell distribution function of electrons,
$$
f_{eq}(\mathbf{r},p,t)=
n(\mathbf{r},t)\Big(\dfrac{m}{2\pi \varkappa T}\Big)^{3/2}
\exp\Big(-\dfrac{p^2}{2\varkappa mT}\Big),
$$
or
$$
f_{eq}(\mathbf{r},p,t)=\dfrac{n(\mathbf{r},t)m^3}{\pi^{3/2}p_T^3}
\exp\Big(-\dfrac{p^2}{p_T^2}\Big),
$$
where $n(\mathbf{r},t)$ is the number density (concentration)
of electrons, $\varkappa$ is the Boltzmann's constant,
$m$ is the electron mass, $\mathbf{p}=m\mathbf{v}$
is the electron momentum,  $p_T=mv_T$ is the thermal momentum of
electrons, $v_T=\dfrac{1}{\sqrt{\beta}}$ is the thermal electron
velocity, $\beta=\dfrac{m}{2\varkappa T}$, $W[f]$ is the Wigner -- Vlasov
functional,
$$
W[f]=\dfrac{1}{(2\pi)^3}\int \Big[U(\mathbf{r}-\dfrac{\hbar
\mathbf{b}}{2},t)\Big]-U(\mathbf{r}+\dfrac{\hbar
\mathbf{b}}{2},t)\Big]\times $$$$ \times f(\mathbf{r},\mathbf{p'},t)
e^{i\mathbf{b}(\mathbf{p'}-\mathbf{p})}\,d^3bd^3p'.
\eqno{(2.2)}
$$

The Wigner function is an analogue of function of distribution for quantum
systems. It is widely used in the diversified questions of
physics. Wigner's function was investigated, for example,
in the works \cite {Arnold} and \cite {Kozlov}.

Let's consider, that distribution electron function depends on one
spatial coordinate $x$, time $t$ and momentum $\mathbf {p}$,
and the electric scalar potential depends on one spatial coordinate
$x$ and time $t$. Then we can write down the equations (2.1) and
(2.2) in the form
$$
\frac{\partial f}{\partial t}+{v_x}\frac{\partial f}{\partial
x}=\dfrac{ie}{\hbar}W[f]+\nu[f_{eq}(x,p,t)-f(x,\mathbf{p},t)],
\eqno{(2.3)}
$$

$$
W[f]=\dfrac{1}{(2\pi)^3}\int \Big[U(x-\dfrac{\hbar
b_x}{2},t)\Big]-U(x+\dfrac{\hbar
{b_x}}{2},t)\Big] \times $$$$ \times f(x,\mathbf{p'},t)
e^{i\mathbf{b}(\mathbf{p'}-\mathbf{p})}\,d^3bd^3p'.
\eqno{(2.4)}
$$

We will linearize the locally equilibrium function $f_{eq}$ in
terms of absolute Maxwell's distribution
$$
f_M(c)=n_0\Big(\dfrac{\beta}{\pi}\Big)^{3/2}e^{-\beta v^2}
\equiv \dfrac{n_0}{\pi^{3/2}v_T^3}\exp\big(-c^2\big),
$$
where $c$ is the module of dimensionless electron velocity $\mathbf{c}
=\dfrac{\mathbf{v}}{v_T}$.

We take the scalar potential in the form
$$
U(x,t)=U_0e^{i(kx-\omega t)}.
\eqno{(2.5)}
$$

Let's search the electron distribution  function  in the following form:
$$
f=f_M(p)\Big[1+U(x,t)h(\mathbf{p})\Big].
\eqno{(2.6)}
$$

Linearization of $f_{eq}$ leads us to expression
$$
f_{eq}=f_M(p)\Big(1+\dfrac{n_1(x,t)}{n_0}\Big),
$$
where
$$
n_1(x,t)\equiv\delta n(x,t)=n(x,t)-n_0.
$$

From the law of conservation of number of particles
$$
\int (f_{eq}-f)d^3v=0
$$
we find that
$$
\dfrac{n_1(x,t)}{n_0}=U(x,t)\dfrac{1}{\sqrt{\pi}}
\int\limits_{-\infty}^{\infty}e^{-\mu^2}h(\mu)d\mu, \quad
\mu=c_x=\dfrac{p_x}{p_T}.
\eqno{(2.7)}
$$

Now we can write down the equation (2.3) in the following form
$$
f_M(v)h(v_x)\Big[\nu+i(kv_x-\omega)\Big]U(x,t)=$$$$=
\dfrac{ie}{\hbar}W[f]+\nu U(x,t)f_M(v)A,
\eqno{(2.8)}
$$
where
$$
A=\dfrac{1}{\sqrt{\pi}}
\int\limits_{-\infty}^{\infty}e^{-\mu^2}h(\mu)d\mu.
\eqno{(2.9)}
$$

The expression (2.4) in linear approximation has the form
$$
W[f]=\dfrac{1}{(2\pi)^3}\int \Big[U(x-\dfrac{\hbar
b_x}{2},t)\Big]-U(x+\dfrac{\hbar
{b_x}}{2},t)\Big] \times $$$$ \times f_M({p'})
e^{i\mathbf{b}(\mathbf{p'}-\mathbf{p})}\,d^3bd^3p'.
\eqno{(2.10)}
$$

Now we receive for potential (2.5)
$$
U(x-\dfrac{\hbar b_x}{2},t)-U(x+\dfrac{\hbar
b_x}{2},t)=$$$$=
U_0e^{i(kx-\omega t)}\Big[\exp(-i\dfrac{k\hbar b_x}{2})-
\exp(i \dfrac{k\hbar b_x}{2})\Big].
$$

Let's calculate internal integral in (2.2).
Considering the last equality, we will integrate in (2.10) by $d^3b $.
We have
$$
\dfrac{1}{(2\pi)^3}\int \Big[\exp(-i\dfrac{k\hbar b_x}{2})-
\exp(i \dfrac{k\hbar b_x}{2})\Big]\times $$$$ \times
e^{ib_x(p'_x-p_x)}
e^{ib_y(p'_y-p_y)}e^{ib_z(p'_z-p_z)}db_xdb_ydb_z=
$$
$$
=\delta(p_y-p_y')\delta(p_z-p_z')
\Big[\delta\Big(p_x-p_x'+\dfrac{\hbar k}{2}\Big)-
\delta\Big(p_x-p_x'-\dfrac{\hbar k}{2}\Big)\Big].
$$

Substituting this equality in (2.10), we receive, that
$$
W[f]=U(x,t)\int \delta(p_y-p_y')\delta(p_z-p_z')\times $$$$
\times
\Big[\delta\Big(p_x-p_x'+\dfrac{\hbar k}{2}\Big)-
\delta\Big(p_x-p_x'-\dfrac{\hbar k}{2}\Big)\Big]
f_M({p}')dp_x'dp_y'dp_z'.
\eqno{(2.11)}
$$

It is necessary to us to integrate by momentums.
Let's notice, that
$$
\int\limits_{-\infty}^{\infty}
\delta(p_y-p_y')e^{-{p_y'}^2/p_T^2}dp_y'=e^{-p_y^2/p_T^2}=e^{-c_y^2},
$$
$$
\int\limits_{-\infty}^{\infty}
\delta(p_z-p_z')e^{-{p_z'}^2/p_T^2}dp_y'=e^{-p_z^2/p_T^2}=e^{-c_z^2},
$$
$$
\int\limits_{-\infty}^{\infty}\delta\Big(p_x-p_x'\pm
\dfrac{\hbar k}{2}\Big)e^{-{p_x'}^2/p_T^2}\,dp_x'=$$$$=
\exp\Big(-\dfrac{\Big(p_x\pm\frac{\hbar k}{2}\Big)^2}{p_T^2}\Big)=
\exp\Big(-\Big(\mu\pm \frac{\hbar k}{2m v_T}\Big)^2\Big)=
e^{-(\mu\pm q/2)^2},
$$
where $\mu=c_x$, $q=k/k_T$, $k_T=mv_T/\hbar$ is the thermal wave
electron number.

According to (2.11)  we  receive further
$$
W[f]=U(x,t)\Big[f_M^+-f_M^-\Big],
\eqno{(2.12)}
$$
where
$$
f_M^{\pm}\equiv f_M^{\pm}(\mathbf{c})=\dfrac{n_0}{\pi^{3/2}v_T^3}
\exp\Big[-(\mu\pm q/2)^2-c_y^2-c_z^2\Big].
$$

By means of (2.11) we will rewrite the equation (2.8) in the form
$$
h(\mu)\Big(1-i\omega\tau+ik_1\mu\Big)=
\dfrac{ie}{\hbar \nu}\dfrac{f^+_M-f^-_M}{f_M}+A,
\eqno{(2.13)}
$$
where $k_1$ is the dimensionless wave number, $k_1=kl,l=v_T\tau$
is the mean free path of electrons.

We receive now from the equation (2.13)
$$
e^{-\mu^2}h(\mu)=\dfrac{ie}{\hbar \nu}\cdot
\dfrac{e^{-(\mu+q/2)^2}-e^{-(\mu-q/2)^2}}
{1-i\omega\tau+ik_1\mu}+
\dfrac{A\,e^{-\mu^2}}{1-i\omega\tau+ik_1\mu}.
\eqno{(2.14)}
$$

For finding of the constant $A$ we will substitute (2.14) in (2.9).
As a result we receive
$$
A=-\dfrac{ie}{\hbar \nu}
\dfrac{J_1(\omega,k,q)}{1-T_0(\omega,k)},
\eqno{(2.15)}
$$
where
$$
T_0=T_0(\omega,k)=\dfrac{1}{\sqrt{\pi}}\int\limits_{-\infty}^{\infty}
\dfrac{e^{-\mu^2}\,d\mu}{1-i\omega\tau+ik_1\mu},
$$
$$
J_1=J_1(\omega,k,q)=\dfrac{1}{\sqrt{\pi}}\int\limits_{-\infty}^{\infty}
\dfrac{e^{-(\mu-q/2)^2}-e^{-(\mu+q/2)^2}}
{1-i\omega\tau+ik_1\mu}\,d\mu.
$$

\begin{center}
\bf 3. Conductivity and permettivity
\end{center}

We consider the connection between electric field and potential
$$
\mathbf{E}(x,t)=-{\rm grad}\;\mathbf{E}(x,t).
$$
Therefore
$$
E_x(x,t)=-\dfrac{\partial U(x,t)}{\partial t}=-ikU(x,t).
$$

From the definition of the longitudinal electric conductivity
$j_x(x,t)=\sigma_l E_x(x,t)$, we find that
$j_x(x,t)=-ik\sigma_lU(x,t)$, hence $\dfrac{\partial j_x}{\partial x}=
\sigma_lk^2U(x,t)$.

From the equation of  continuity for current and charge
densities
$$
\dfrac{\partial \rho}{\partial t}+
\dfrac{\partial j_x}{\partial x}=0
$$
according to the last equality we receive that
$$
\dfrac{\partial \rho}{\partial t}=-\dfrac{\partial j_x}{\partial
x}=
-\sigma_l k^2U(x,t).
$$

On the other hand, from the definition  of a charge of a density
$$
\rho=e\int fd\Omega=
e\int f_M(v)[1+U(x,t)h(v_x)]\,d^3v,
$$
we find
$$
\dfrac{\partial \rho}{\partial t}=-i\omega eU(x,t)\int
f_M(v)h(v_x) d^3v.
$$

Now we receive the relation for longitudinal electroconductivity
$$
\sigma_l=\dfrac{ie\omega}{k^2}\int f_M h(v_x)\,d^3v=
\dfrac{ie\omega n_0}{k^2}\dfrac{1}{\sqrt{\pi}}
\int\limits_{-\infty}^{\infty}e^{-\mu^2}h(\mu)d\mu=
\dfrac{ie\omega n_0}{k^2}A.
$$

Substituting the expression (2.15) instead of $A$ , we receive the
expression for
conductivity of quantum plasma
$$
\sigma_l=\dfrac{e^2n_0\omega
\tau}{k^2 \hbar}\dfrac{J_1}{1-T_0}=\sigma_0\dfrac{q\omega}{k_1\nu}
\dfrac{J_1}{1-T_0}.
\eqno{(3.1)}
$$

Let's transform integrals $T_0$ and $J_1$. We receive, that
$$
T_0=\dfrac{1}{ik\tau v_T}\cdot\dfrac{1}{\sqrt{\pi}}
\int\limits_{-\infty}^{\infty}\dfrac{e^{-\mu^2\,d\mu}}
{\mu-\dfrac{\omega -i \nu}{kv_T}}=-\dfrac{i \nu}{kv_T}t(z)=
-\dfrac{i}{k_1}t(z),
$$
where
$$
t(z)=\dfrac{1}{\sqrt{\pi}}\int\limits_{-\infty}^{\infty}
\dfrac{e^{-\mu^2}\,d\mu}{\mu-z}, \qquad z=\dfrac{\omega+ i \nu}{
kv_T}.
$$

Integral $J_1$ we will present in the form
$$
J_1=\dfrac{1}{ik\tau v_T}J(z,q)=
-\dfrac{i \nu}{kv_T}J(z,q)=-\dfrac{i}{k_1}J(z,q),
$$
where
$$
J(z,q)=\dfrac{1}{\sqrt{\pi}}
\int\limits_{-\infty}^{\infty}\dfrac{e^{-(\mu-q/2)^2}-
e^{-(\mu+q/2)^2}}
{\mu-z}d\mu.
$$

Hence, electric conductivity of quantum plasma
according to (3.1)  is equal to
$$
\sigma_l=-i\sigma_0\dfrac{q\omega}{k_1^2\nu}
\dfrac{J(z,q)}{1+it(z)/k_1},\qquad z=\dfrac{\omega
+i \nu}{kv_T}.
\eqno{(3.2)}
$$

Dielectric permettivity of plasma we will find, if
we use (3.1) or (3.2). According to definition
$$
\varepsilon_l=1+\dfrac{4\pi i}{\omega}\sigma_l
$$
we find accordingly
$$
\varepsilon_l=1+i\dfrac{\omega_p^2q}{\nu^2k_1}
\dfrac{J_1}{1-T_0}.
\eqno{(3.3)}
$$
or
$$
\varepsilon_l=1+\dfrac{\omega_p^2q}{\nu^2k_1^2}
\dfrac{J(z,q)}{1+it(z)/k_1}.
\eqno{(3.4)}
$$

Here $\omega_p$ is the own plasma (Langmuir) frequency
$$
\omega_p^2=\dfrac{4\pi e^2 n_0}{m}.
$$

Let's show that in a limit when $\hbar\to 0$, the expression
$\varepsilon_l$ for longitudinal permettivity of quantum
plasmas passes into corresponding expression for
classical plasma
$$
\varepsilon_l^\circ=1+
\dfrac{2\omega_p^2}{k^2v_T^2}\dfrac{\lambda_0(z)}
{1-T_0},
$$
where $\lambda_0(z)$ is the well known (see \cite{Kampen})
dispersion plasma function entered by Van Kampen
$$
\lambda_0(z)=\dfrac{1}{\sqrt{\pi}}\int\limits_{-\infty}^{\infty}
\dfrac{e^{-\mu^2}\mu\,d\mu}
{\mu-\dfrac{\omega+i \nu}{kv_T}}.
$$

We notice that
$$
\lim\limits_{\hbar \to 0}
\dfrac{e^{-(\mu-q/2)^2}-e^{-(\mu+q/2)^2}}{\hbar}=
\dfrac{2ke^{-\mu^2}}{mv_T}.
$$

By means of this relationship we receive
$$
\lim\limits_{\hbar \to 0}\dfrac{J}{\hbar}=\dfrac{2k}{mv_T}
\lambda_0(z).
$$

Thus, passing to a limit at $\hbar\to 0$ in the
expression (3.4), we receive in accuracy the relationship for
conductivity of classical plasma.

We will transform a denominator from formulas (3.3)
$$
1-T_0=1+\dfrac{i \nu}{kv_T}\dfrac{1}{\sqrt{\pi}}
\int\limits_{-\infty}^{\infty}\dfrac{e^{-\mu^2}\,d \mu}
{\mu-\dfrac{\omega+i \nu}{kv_T}}=1+\dfrac{i \nu}{kv_T}t(z).
$$

We notice that
$
\lambda_0(z)=1+z\;t(z).
$
Further we receive
$$
1-T_0=1+\dfrac{i \nu}{\omega+i \nu}\cdot \dfrac{\omega+i \nu}
{kv_T}t(z)=1+\dfrac{i \nu}{\omega+i \nu}zt(z)=
$$
$$
=\dfrac{\omega+i \nu+i \nu zt(z)}{\omega+i \nu }=
\dfrac{\omega+i \nu \lambda_0(z)}{\omega+i \nu}.
$$

Taking into account this identity dielectric permettivity of the quantum
and classical plasma accordingly equals
$$
\varepsilon_l=1+\dfrac{\omega_p^2z^2}{(\omega+i \nu)q}
\cdot\dfrac{t(z-q/2)-t(z+q/2)}{\omega+i \nu\lambda_0(z)}
\eqno{(3.5)}
$$
and
$$
\varepsilon_l^\circ=1+\dfrac{2\omega_p^2}{k^2v_T^2}\cdot
\dfrac{(\omega+i \nu)\lambda_0(z)}{\omega+i \nu\lambda_0(z)}.
\eqno{(3.6)}
$$

The expression (3.6) can be written down in the form
$$
\varepsilon_l^\circ=1+\dfrac{2\omega_p^2}{\omega+i \nu}\cdot
\dfrac{z^2\lambda_0(z)}{\omega+i \nu
\lambda_0(z)}.
\eqno{(3.6')}
$$

The formula (3.5) can be presented in one of the equivalent
forms
$$
\varepsilon_l=1+\dfrac{\omega_p^2 }{(\omega+i \nu)(
\omega+i \nu \lambda_0(z))}\cdot\dfrac{z^2[t(z-q/2)-
t(z+q/2)]}{q},
\eqno{(3.7)}
$$
or
$$
\varepsilon_l=1+\dfrac{\omega_p^2 (\omega+i \nu)}
{k^2v_T^2(\omega+i \nu \lambda_0(z))}\cdot\dfrac{t(z-q/2)-
t(z+q/2)}{q},
\eqno{(3.7')}
$$

We present the formula (3.7) in the form
$$
\varepsilon_l=1+\dfrac{\omega_p^2 }{(\omega+i \nu)(
\omega+i \nu \lambda_0(z))}\cdot\dfrac{f(z,q)}{q},
\eqno{(3.8)}
$$
Here
$$
f(z,q)=z^2[t(z-q/2)-t(z+q/2)].
$$

We notice that
$$
\dfrac{t(z-q/2)-t(z+q/2)}{q}=-\dfrac{1}{\sqrt{\pi}}
\int\limits_{-\infty}^{\infty}
\dfrac{e^{-\mu^2}\,d\mu}{(\mu-z)^2-q^2/4}.
$$

We will designate
$$
J_0(z)=-\dfrac{1}{\sqrt{\pi}}
\int\limits_{-\infty}^{\infty}
\dfrac{e^{-\mu^2}\,d\mu}{(\mu-z)^2-q^2/4}.
$$

By means of this designation we will present the formula (3.8) in
the form
$$
\varepsilon_l=1+\dfrac{\omega_p^2 z^2 }{(\omega+i \nu)(
\omega+i \nu \lambda_0(z))}\cdot J_0(z),
\eqno{(3.9)}
$$
or
$$
\varepsilon_l=1+\dfrac{\omega_p^2}{k^2v_T^2}\cdot
\dfrac{(\omega+i \nu)J_0(z)}{\omega+i \nu \lambda_0(z)}, \qquad
z=\dfrac{\omega+i \nu}{kv_T}.
\eqno{(3.9')}
$$

Further we will enter dimensionless parametres
$$
x=\dfrac{\omega}{k_Tv_T}=\dfrac{\omega}{\nu k_Tl}=
\dfrac{\omega}{\nu k_{01}}, \qquad
y=\dfrac{\nu}{k_Tv_T}=\dfrac{1}{k_Tl}=\dfrac{1}{k_{01}},
$$
where $k_{01}=k_Tl$ is the dimensionless wave thermal number of electrons.

We will present the formula $(3.9')$ in the explicit form using
the entered dimensionless parametres
$$
\varepsilon_l=1-\dfrac{x_p^2}{q^2}
\dfrac{\dfrac{x+iy}{\sqrt{\pi}}\displaystyle\int\limits_{-\infty}^{\infty}
\dfrac{e^{-\tau^2}d\tau}{[\tau-(x+iy)/q]^2-q^2/4}}
{x+\dfrac{iy}{\sqrt{\pi}}\displaystyle
\int\limits_{-\infty}^{\infty}\dfrac{\tau e^{-\tau^2}d\tau}
{\tau-(x+iy)/q}},
\eqno{(3.10)}
$$
where $x_p$ is the dimensionless plasma frequency,
$x_p=\dfrac{\omega_p}{k_Tv_T}.$

\begin{center}
\bf 4.  Dielectric permettivity properties
\end{center}

Let's find a long-wave limit (at $k\to 0$) of dielectric
permettivity (3.9). We will present this formula in the form
$$
\varepsilon_l=1-\dfrac{\omega_p^2}{(\omega+i \nu)(
\omega+i \nu \lambda_0(z))}\cdot \dfrac{z^2}{\sqrt{\pi}}
\int\limits_{-\infty}^{\infty}\dfrac{e^{-\mu^2}\,d\mu}
{(\mu-z)^2-q^2/4},
$$

We will transform the last factor as follows
$$
\dfrac{z^2}{\sqrt{\pi}}
\int\limits_{-\infty}^{\infty}\dfrac{e^{-\mu^2}\,d\mu}
{(\mu-z)^2-q^2/4}=\dfrac{1}{\sqrt{\pi}}
\int\limits_{-\infty}^{\infty}\dfrac{e^{-\mu^2}\,d\mu}
{1-\dfrac{2\mu}{z}-\dfrac{q^2/4-\mu^2}{z^2}}.
$$

From here it is easy to notice, that at $z\to \infty$ the last
integral tends to unit. It means, that
$$
\varepsilon_l(\omega,\nu,k=0)
=1-\dfrac{\omega_p^2}{(\omega+i \nu)(\omega+i \nu \lambda_0(z))}.
\eqno{(4.1)}
$$

From the expression (4.1) we can see that at $\nu=0$ we receive
classical result for collisionless plasma
$$
\varepsilon_l(\omega,0,0)=1-\dfrac{\omega_p^2}{\omega^2}.
$$

Let's consider special cases of dielectric permettivity.

In the formulas (3.7) or (3.9) we will put $\nu=0$. Then for the
dielectric permettivity in collisionless plasma we receive
$$
\varepsilon_l(\omega,k,\nu=0)=1-\dfrac{\omega_p^2}{k^2v_T^2\sqrt{\pi}}
\int\limits_{-\infty}^{\infty}\dfrac{e^{-\tau^2}\;d\tau}
{\Big(\tau-\dfrac{\omega}{kv_T}\Big)^2-\Big(\dfrac{\hbar
k}{2mv_T}\Big)^2}.
\eqno{(4.2)}
$$

We present the formula (4.2) in dimensionless parameters
$$
\varepsilon_l(x,q)=1-\dfrac{x_p^2}{q^2\sqrt{\pi}}
\int\limits_{-\infty}^{\infty}
\dfrac{e^{-\tau^2}\;d\tau}{(\tau-x/q)^2-q^2/4},
\eqno{(4.3)}
$$
or
$$
\varepsilon_l(x,q)=1+\dfrac{x_p^2}{q^3\sqrt{\pi}}
\Bigg[\int\limits_{-\infty}^{\infty}\dfrac{e^{-\tau^2}d\tau}{\tau-x/q+q/2}-
\int\limits_{-\infty}^{\infty}\dfrac{e^{-\tau^2}d\tau}{\tau-x/q-q/2}\Bigg].
\eqno{(4.4)}
$$

Now we will consider a low-frequency limit ($\omega\to 0$) in
collisional  form ($ \nu \ne 0$). In this limit we receive
$$
\varepsilon_l(\omega=0,k,\nu)=
1+\dfrac{\omega_p^2}{k^2v_T^2}\dfrac{J_0(z)}{\lambda_0(z)}, \qquad
z=\dfrac{i \nu }{kv_T}.
\eqno{(4.5)}
$$

We can rewrite down the formula (4.5) in the  form
$$
\varepsilon_l(0,y,q)=1+\dfrac{x_p^2}{q^2}\dfrac{J_0(iy/q)}{\lambda_0(iy/q)}.
\eqno{(4.6)}
$$

We transform the formula (4.6) into the following explicit form
$$
\varepsilon_l(0,y,q)=1-\dfrac{x_p^2}{q^3}\cdot
\dfrac{\displaystyle \dfrac{1}{\sqrt{\pi}}\int\limits_{-\infty}^{\infty}
\dfrac{e^{-\tau^2}d\tau}
{(\tau-iy/q)^2-q^2/4}}{\displaystyle 1+\dfrac{iy}{q\sqrt{\pi}}
 \int\limits_{-\infty}^{\infty}
\dfrac{e^{-\tau^2}\,d\tau}{\tau-iy/q}}.
\eqno{(4.7)}
$$

\begin{center}
\bf 5. Plasma oscillations in long--wave limit
\end{center}

Let's consider the linearization of the dielectric permettivity by
quantum parameter $q=\dfrac{\hbar k}{mv_T}=\dfrac{k}{k_T}$ for
the case when $|z|\gg 1$, and $q\ll 1$, where
$z=\dfrac{\omega +i\nu}{kv_T}$.

We will expand the function
$$
f(z,q)=z^2[t(z-q/2)-t(z+q/2)].
$$
at large $z$ and small $q$
$$
\dfrac{f(z,q)}{q}=-1-\dfrac{3}{2z^2}-\dfrac{q^2}{4z^2}-\dfrac{15}{4z^4}.
$$

We consider the integral in the complex plane
$$
J(z,q)=\dfrac{1}{\sqrt{\pi}}\int\limits_{-\infty}^{\infty}
\dfrac{e^{-(\mu-q/2)^2}-e^{-(\mu+q/2)^2}}{\mu-z}d\mu.
\eqno{(5.1)}
$$

We will present the subintegral function in the form
$$
\dfrac{e^{-(\mu-q/2)^2}-e^{-(\mu+q/2)^2}}{\mu-z}=
\dfrac{e^{-\mu^2-q^2/4}}{\mu-z}
\Big[e^{q\mu}-e^{-q\mu}\Big]=
$$
$$
=2q\mu \Big[1+\dfrac{1}{6}(q\mu)^2\Big]
\dfrac{e^{-\mu^2-q^2/4}}{\mu-z}.
$$

Hence, we can present the integral (5.1) in the form
$$
\dfrac{J}{2q}=\dfrac{e^{-q^2/4}}{\sqrt{\pi}}
\int\limits_{-\infty}^{\infty}\dfrac{e^{-\mu^2}\mu\,d\mu}{\mu-z}+
\dfrac{q^2e^{-q^2/4}}{6\sqrt{\pi}}
\int\limits_{-\infty}^{\infty}
\dfrac{e^{-\mu^2}\mu^3\,d\mu}{\mu-z}.
\eqno{(5.2)}
$$

We will notice, that
$$
\dfrac{1}{\sqrt{\pi}}
\int\limits_{-\infty}^{\infty}\dfrac{e^{-\mu^2}\mu^3\,d\mu}{\mu-z}=
\dfrac{1}{\sqrt{\pi}}
\int\limits_{-\infty}^{\infty}\dfrac{e^{-\mu^2}
\mu(\mu^2-z^2+z^2)\,d\mu}{\mu-z}=
$$
$$
=\dfrac{1}{\sqrt{\pi}}
\int\limits_{-\infty}^{\infty}e^{-\mu^2}\Big[\mu(\mu+z)+
z^2\dfrac{\mu}{\mu-z}\Big]d\mu=\dfrac{1}{2}+z^2\lambda_0(z).
$$

Thus, the expression (5.2) can be written down in the following
manner
$$
\dfrac{J}{2q}=e^{-q^2/4}
\Big\{\lambda_0(z)+\dfrac{q^2}{6}
\Big[\dfrac{1}{2}+z^2\lambda_0(z)\Big]\Big\}.
\eqno{(5.3)}
$$

Now we will return to the dielectric permettivity (3.9) and with
the help of (5.3) we will present it in the form
$$
\varepsilon_l=1+\dfrac{2\omega_p^2 z^2}{(\omega+i \nu)
(\omega+i \nu \lambda_0(z))}\cdot
e^{-q^2/4}
\Big\{\lambda_0(z)+\dfrac{q^2}{6}
\Big[\dfrac{1}{2}+z^2\lambda_0(z)\Big]\Big\}.
\eqno{(5.4)}
$$

In the expansion (5.4) by degrees $k^2$ we will keep
members with degrees not above than $k^4$, taking into
account the equalities proportional to $k$
$$
q=\dfrac{\hbar }{mv_T}\cdot k\quad \text{and}\quad
\dfrac{1}{z}=\dfrac{v_T}{\omega+i \nu}\cdot k.
$$

We will search for the solutiom of the dispersion equation
$$
\varepsilon_l(\omega,k)=0
\eqno{(5.5)}
$$
for small  $k$, i.e. we seek the solution $\omega=\omega(k)$
of the equation (5.5) for the case when
$$
|z|=\left|\dfrac{\omega+i \nu}{kv_T}\right|\gg 1, \qquad
\nu \ll \omega.
\eqno{(5.6)}
$$

We will use the expansions
$$
\lambda_0(z)=-\dfrac{1}{2z^2}-\dfrac{3}{4z^4}-
\dfrac{15}{8z^6}-\cdots,\qquad z\to \infty,
$$
$$
\dfrac{1}{2}+z^2\lambda_0(z)=-\dfrac{3}{4z^2}-\dfrac{15}{8z^4}.
$$

By means of these expansions we receive
$$
2z^2e^{-q^2/4}
\Big\{\lambda_0(z)+\dfrac{q^2}{6}
\Big[\dfrac{1}{2}+z^2\lambda_0(z)\Big]\Big\}=
$$
$$
=-\Big\{1+\dfrac{3}{2z^2}+\dfrac{15}{4z^4}
+\dfrac{q^2}{4z^2}\Big\}.
$$

Let's rewrite the equation (5.5) in the explicit form
$$
(\omega+i \nu)(\omega+i \nu \lambda_0(z))-
$$
$$
-\omega_p^2 \Big(1+\dfrac{3}{2z^2}+\dfrac{q^2}
{4z^2}+\dfrac{15}{4z^4}-\dfrac{q^4}{16}\Big)=0,
$$
or, taking into account (5.6), we have
$$
\omega^2-\omega_p^2 \Big(1+\dfrac{3}{2z^2}+\dfrac{q^2}
{4z^2}+\dfrac{15}{4z^4}\Big)=0,
\eqno{(5.7)}
$$

Let's consider the real part of the equation (5.7) at
conditions (5.6) near to plasma resonance, i.e. we will
search for the solution (5.7) in the form
$$
\omega_k^2=\omega_p^2(1+\varepsilon), \qquad
|\varepsilon|\ll 1.
\eqno{(5.8)}
$$

Substituting (5.8) in the equation (5.7), we come to the
following equation
$$
\varepsilon=\dfrac{3}{2z^2}+\dfrac{q^2}{4z^2}+\dfrac{15}{4z^4}.
$$
Let's write down this equation in the explicit form
$$
\varepsilon=\dfrac{3k^2v_T^2(1-\varepsilon)}{2\omega_p^2}+
\dfrac{\hbar^2 k^4v_T^2(1-\varepsilon)}{4m^2v_T^2\omega_p^2}+
\dfrac{15k^4v_T^4(1-2\varepsilon)}{4\omega_p^4}.
$$

From here we receive
$$
\varepsilon=\dfrac{3k^2}{k_D^2}+\dfrac{6k^4}{k_D^4}+
\dfrac{\hbar^2k^4}{2m^2v_T^2k_D^2},
$$
where $k_D$ is the Debye wave number inverse proportional to the
Debue radius $r_D$,
$$
k_D=\dfrac{\sqrt{2}\omega_p}{v_T}, \qquad
k_D=\dfrac{1}{r_D}.
$$

According to (5.8) we have found the dependence
$$
\omega^2(k)=\omega_p^2\Big(1+\dfrac{3k^2}{k_D^2}+\dfrac{6k^4}{k_D^4}
+\dfrac{\hbar^2k^4}{2m^2v_T^2k_D^2}\Big).
\eqno{(5.9)}
$$

Let's rewrite (5.9) in two equivalent forms
$$
\omega^2(k)=\omega_p^2\Big(1+\dfrac{3\varkappa T}{m\omega_p^2}k^2+
\dfrac{6\varkappa^2 T^2}{m^2\omega_p^4}k^4+
\dfrac{\hbar^2k^4}{4m^2\omega_p^2}\Big),
$$
or
$$
\omega^2(k)=\omega_p^2+\dfrac{3\varkappa T}{m}k^2+
\dfrac{6\varkappa^2T^2}{m^2\omega_p^2}k^4+\dfrac{\hbar^2 k^4}
{4m^2}.
\eqno{(5.10)}
$$

$$
\omega^2(k)=\omega_p^2+\dfrac{3\varkappa T}{m}k^2+
\dfrac{6\varkappa^2T^2}{m^2\omega_p^2}k^4\Big(1+\dfrac{Q^2}
{24}\Big), \quad Q=\dfrac{\hbar \omega_p}{\varkappa T}.
\eqno{(5.10a)}
$$

Here $Q$ is the quantum parameter, showing how much essential
quantum amendments to quantity of frequency of plasma oscillations are.

With use of thermal electron velocity $v_T $ the formula
(5.10) can be written down in the form
$$
\omega^2(k)=\omega_p^2+\dfrac{3}{2}\dfrac{v_T^2}{\omega_p^2}k^2+
\dfrac{3}{2}\dfrac{v_T^4}{\omega_p^4}k^4+\dfrac{\hbar^2k^4}{4m^2}.
$$

The last member in the formula (5.10) (the quantum amendment) in
accuracy coincides with a corresponding member from the formula
(5.8) from  Manfredi's work \cite {Manf}, deduced for degenerate plasma.

From the expression (5.10) one can see that when $\hbar=0$ it passes
into classical expression:
$$
\omega_k^2=\omega_p^2\Big(1+\dfrac{3k^2}{k_D^2}+\dfrac{6k^4}
{k_D^4}\Big).
\eqno{(5.11)}
$$

Now we will find damping of plasma oscillations in quantum plasma.
Let's present the dispersion equation (5.5) by means of (5.4)
in the form
$$
(\omega+i \nu)(\omega+i \nu \lambda_0(z))+\omega_p^2
e^{-q^2/4}\Bigg[2z^2\lambda_0(z)
\Big[1+\dfrac{(qz)^2}{6}\Big]\Bigg]=0.
\eqno{(5.12)}
$$

Let's calculate decrement of decrease $\gamma_k$.
Let's consider for this purpose frequency
$\omega$ as complex value and assume also, that
$$
\omega=\omega_k+i\gamma_k,  \quad \omega_k=\omega(k),
\quad |\gamma_k|\ll \omega_k.
\eqno{(5.13)}
$$

Let's substitute the expression (5.13) in the equation (5.12)  and we will
mark out in the received equation an imaginary part, using the inequalities
(5.6). As a result we receive the equation
$$
2\omega_k\gamma_k+\omega_k \nu=-2\sqrt{\pi}\omega_p^2
\dfrac{\omega_k^3}{k^3v_T^3}\exp\Big(-\dfrac{\omega_k^2}
{k^2v_T^2}-\alpha^2\Big)
\Big(1+\dfrac{2}{3}(\alpha z)^2\Big),
$$
from which we receive the decrement of damping
$$
\gamma_k=-\dfrac{\nu}{2}-\sqrt{\dfrac{\pi}{8}}\omega_p
\dfrac{k_0^3}{k^3}\exp\Big(-\dfrac{3}{2}-
\dfrac{k_0^2}{2k^2}\Big)(1-\dfrac{q^2}{4})
\Big(1+\dfrac{q^2z^2}{6}\Big).
\eqno{(5.14)}
$$

It is obvious, that at $\hbar=0$ the formula (5.14) passes into
the well known classical result
$$
\gamma_k=-\dfrac{\nu}{2}-\sqrt{\dfrac{\pi}{8}}\omega_p
\dfrac{k_D^3}{k^3}\exp\Big(-\dfrac{3}{2}-\dfrac{k_D^2}{2k^2}\Big).
\eqno{(5.15)}
$$
Believing in (5.15) $\nu=0$, we come to the well known Landau
damping formula
$$
\gamma_k=-\sqrt{\dfrac{\pi}{8}}\omega_p
\dfrac{k_D^3}{k^3}\exp\Big(-\dfrac{3}{2}-\dfrac{k_D^2}{2k^2}\Big).
\eqno{(5.16)}
$$

\begin{center}
\bf 6. Comparison with Mermin's result
\end{center}

Mermin (see Mermin N.D. \cite{Mermin}) has been received
the general expression of dielectric function
$$
\varepsilon^M(\omega,k)=
1+\dfrac{(\omega+i \nu)\Big[\varepsilon^\circ
(\omega+i \nu,k)-1\Big]}
{\omega +i\nu \dfrac{\varepsilon^\circ(\omega+i \nu,k)-1}
{\varepsilon^\circ(0,k)-1}}.
\eqno{(6.1)}
$$

The formula (3.1) is received on the basis of the kinetic equation
for one-partial matrix of density $\rho$
$$
\dfrac{\partial \rho}{\partial t}+i[\E+V,+
\rho]=\dfrac{{\rho}^{\;l.e.}-\rho}{\tau}
$$
with relaxation time $\tau$ and local equilibrium density matrix
$\rho^{\;l.e.}$, and $\E$ is the kinetic energy of
electrons,
$V$ is the self-consistent potential, $[\cdot,\cdot]$ is the commutator,
$\rho^{\;l.e.}$ is the local equilibrium distribution function
of electrons\; (l.e.$\equiv$ local \;equilibrium),
$$
\rho=\dfrac{1}{1+\exp(\varepsilon-\mu-\delta\mu)},
$$
$\mu$ is the dimensionless (normalized) chemical potential.

Dielectric function is calculated in the momentum representation
$$
f(\mathbf{p},\mathbf{q},t)=\Big\langle
\mathbf{p}+\frac{\mathbf{q}}{2}\;\Big|\;\rho\;\Big|\;
\mathbf{p}-\frac{\mathbf{q}}{2}\Big\rangle,
$$
where $\Big|\;\mathbf{p}\Big \rangle$ is the eigen momentum state
$\mathbf{p}$.
For convenience we will transform the density matrix
to Wigner's distribution in phase space
$$
f(\mathbf{p},\mathbf{R},t)=\dfrac{1}{(2\pi)^3}
\int e^{i\mathbf{q}\mathbf{R}}f(\mathbf{p},\mathbf{q},t)d^3q,
$$
where $\mathbf{R}$ is the spatial coordinate.

In the formula (6.1)  the designation $\varepsilon^\circ(\omega,k)$
is entered, it is so-called
Lindhard' dielectric function, i.e.
the dielectric function received for collisionless
plasmas, expression $\varepsilon^\circ(\omega+i\nu,k)$
means, that argument of Lindhard' dielectric function
$\omega$ is replaced formally by $\omega+i\nu$.
According to $(3.7)$
$$
\varepsilon_l^\circ(\omega+i \nu,k)-1=\dfrac{2\omega_p^2}{k^2v_T^2
2q}\Big[t\big(\frac{\omega+i \nu}{kv_T}-\dfrac{q}{2}\big)-
t\big(\frac{\omega+i \nu}{kv_T}+\dfrac{q}{2}\big)\Big],
\eqno{(6.2)}
$$
$$
\varepsilon_l^\circ(0,k)-1=\dfrac{2\omega_p^2}{k^2v_T^2 2q}
\Big[t(-q/2)-t(q/2)\Big].
\eqno{(6.3)}
$$

For comparison we will write out the received dielectric function
for collisional plasma (3.7) in the explicit form
$$
\varepsilon_l(k,\omega)=1+\dfrac{2\omega_p^2(\omega + i \nu)}
{k^2v_T^22q}\cdot
\dfrac{t\big(\frac{\omega+i \nu}{kv_T}-\frac{q}{2}\big)-
t\big(\frac{\omega+i \nu}{kv_T}+\frac{q}{2}\big)}{\omega +i \nu
\lambda_0\big(\frac{\omega+i \nu}{kv_T}\big)}.
\eqno{(6.4)}
$$

From formulas (6.2) and (6.3) we can see that
$$
\dfrac{\varepsilon^\circ(\omega+i
\nu,k)-1}{\varepsilon^\circ(0,k)-1}=\dfrac{t(z-q/2)-t(z+q/2)}
{t(-q/2)-t(q/2)}, \qquad z=\dfrac{\omega+i
\nu}{kv_T}.
\eqno{(6.5)}
$$

By means of (6.2)--(6.5) Mermin's formula (6.1) in our
designations will be written in the form
$$
\varepsilon^M=1+\dfrac{2\omega_p^2(\omega + i \nu)}
{k^2v_T^22q}\cdot \dfrac{t(z-q/2)- t(z+q/2)}{\omega +i \nu
\dfrac{t(z-q/2)-t(z+q/2)}{t(-q/2)-t(q/2)}}.
\eqno{(6.6)}
$$

From (6.4) and (6.6) we can see that the received formula for
dielectric function differs from the corresponding
Mermin' function by that in Mermin's formula the relation
$$
\dfrac{t(z-q/2)-t(z+q/2)}
{t(-q/2)-t(q/2)}, \qquad z=\dfrac{\omega+i\nu}{kv_T}.
$$
is necessary to replace with dispersion plasma function of  Van
Kampen $\lambda_0(z)$.

It is necessary to notice, that at small values of parameter $q$
the Mermin formula is close to true and passes into it when
$\hbar\to 0$ (or $q\to 0$).

Indeed, it is required to show, that the following
limiting transition is satisfied
$$
\lim\limits_{q\to 0}\dfrac{t(z-q/2)-t(z+q/2)}
{t(-q/2)-t(q/2)}=\lambda_0(z).
$$

Let's take advantage of the equalities proved above
$$
\lim\limits_{q\to
0}\dfrac{t(z-q/2)-t(z+q/2)}{2q}=\dfrac{1}{2}
\lim\limits_{q\to 0}J_1(z,q)=\lambda_0(z),
$$
$$
t(-q/2)-t(q/2)=2qe^{-q^2/4}\int\limits_{0}^{q/2}
e^{u^2}\,du.
$$

By means of these equalities we receive that
$$
\lim\limits_{q\to 0}\dfrac{t(z-q/2)-t(z+q/2)}
{t(-q/2)-t(q/2)}=\lim\limits_{q\to 0}
\dfrac{t(z-q/2)-t(z+q/2)}
{2q\big(1-q^2/{6}\big)}=\lambda_0(z).
$$

Thus, at $q\to 0$ (or, that is all the same, $\hbar\to 0$
or $k\to 0$) Mermin's formula and formula deduced in  this work
pass into the same dielectric function
for classical Maxwellian plasma.

\begin{figure}[t]
\begin{center}
\includegraphics[width=0.49\textwidth]{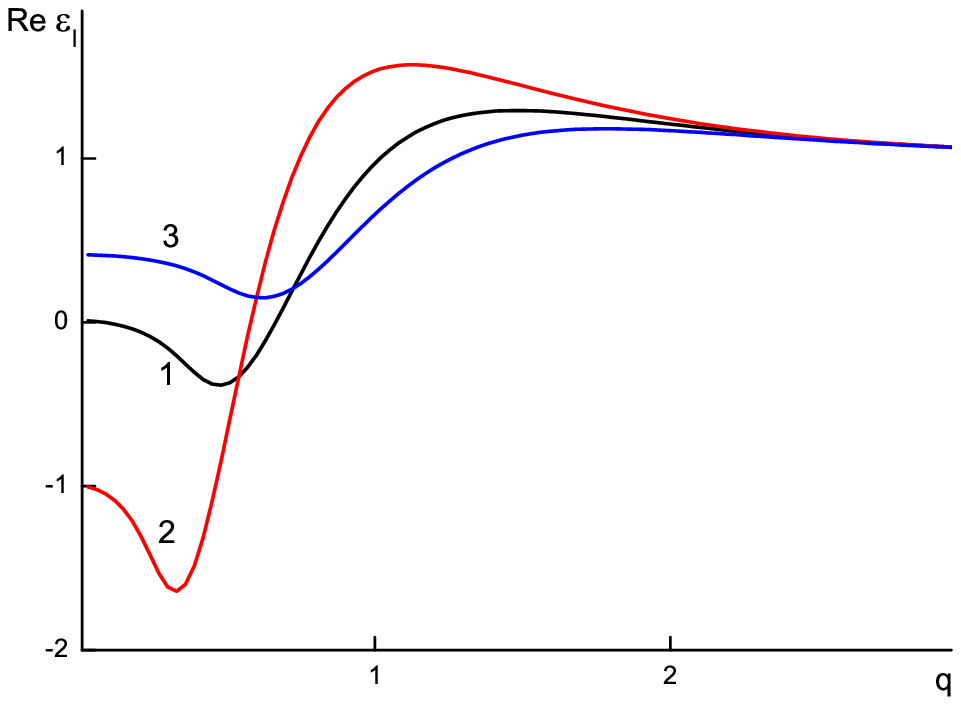}
\hfill
\includegraphics[width=0.49\textwidth]{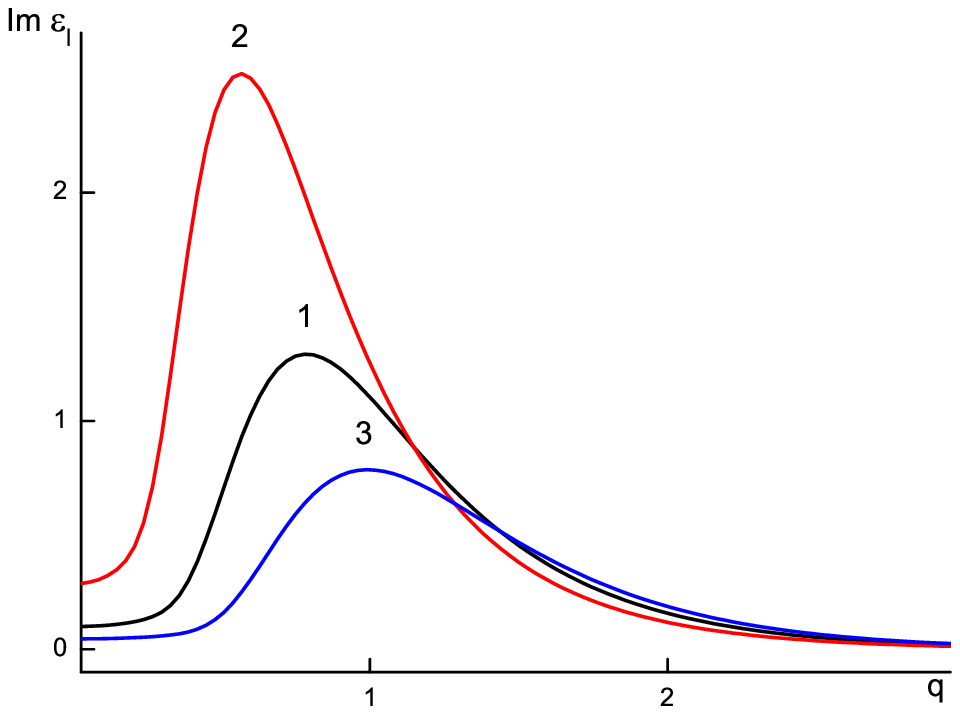}
\\
\parbox[t]{0.47\textwidth}{\hspace{3cm}{Fig. 1.}}
\hfill
\parbox[t]{0.47\textwidth}{\hspace{3cm}{Fig. 2.}}
\end{center}
\end{figure}
\begin{figure}[h]
\begin{center}
\includegraphics[width=0.49\textwidth]{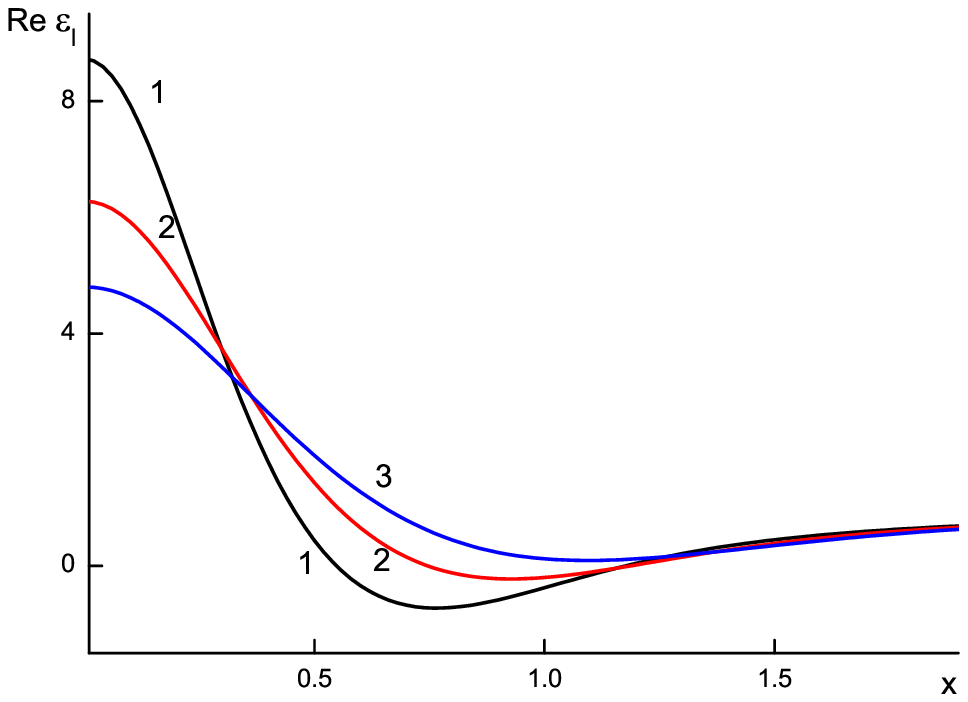}
\hfill
\includegraphics[width=0.49\textwidth]{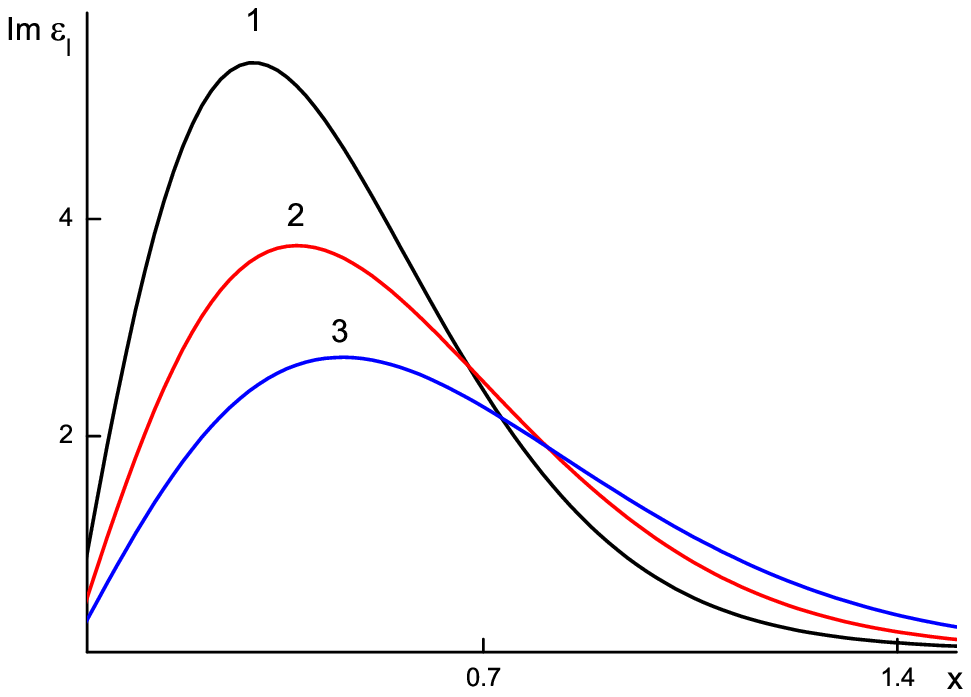}
\\
\parbox[t]{0.47\textwidth}{\hspace{3cm}{Fig. 3.}}
\hfill
\parbox[t]{0.47\textwidth}{\hspace{3cm}{Fig. 4.}}
\end{center}
\end{figure}
\begin{figure}[htb]
\begin{center}
\includegraphics[width=0.49\textwidth]{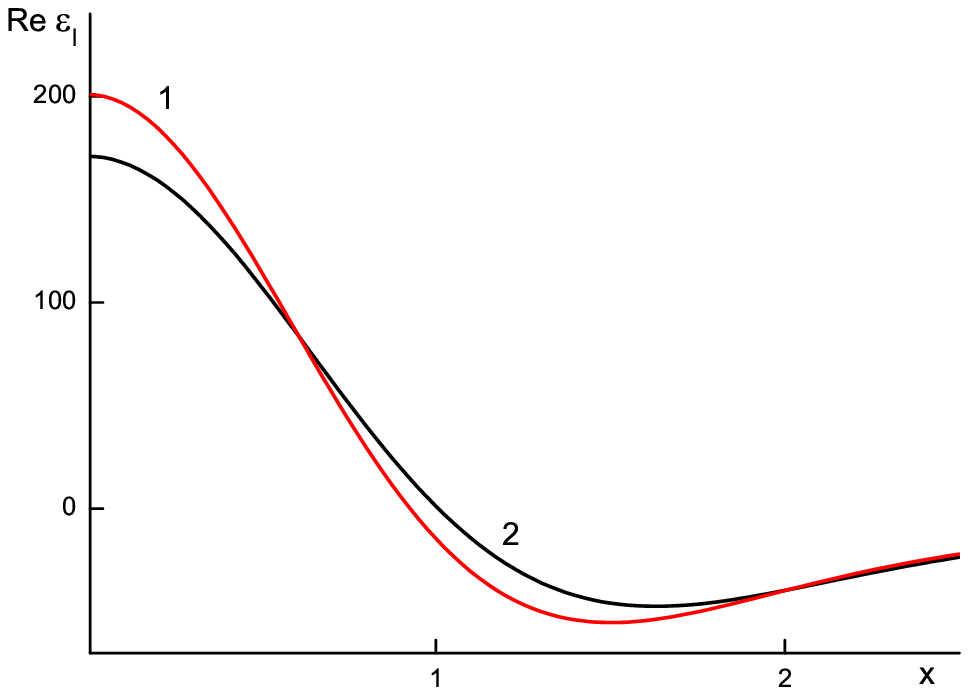}
\hfill
\includegraphics[width=0.49\textwidth]{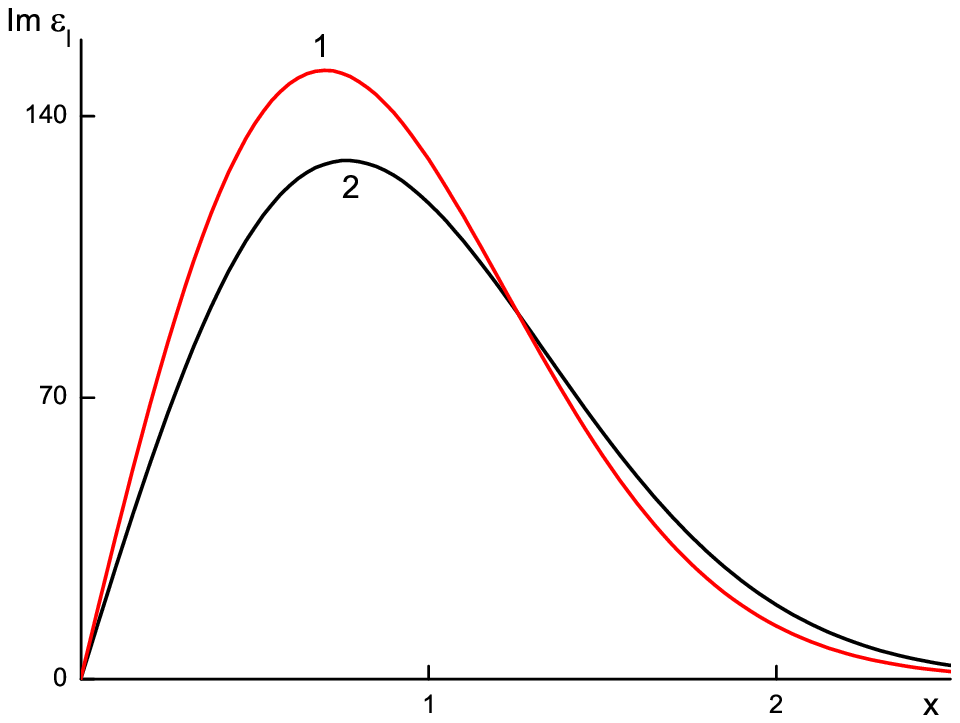}
\\
\parbox[t]{0.47\textwidth}{\hspace{3cm}{Fig. 5.}}
\hfill
\parbox[t]{0.47\textwidth}{\hspace{3cm}{Fig. 6.}}
\end{center}
\end{figure}
\begin{figure}[h]
\begin{center}
\includegraphics[width=0.49\textwidth]{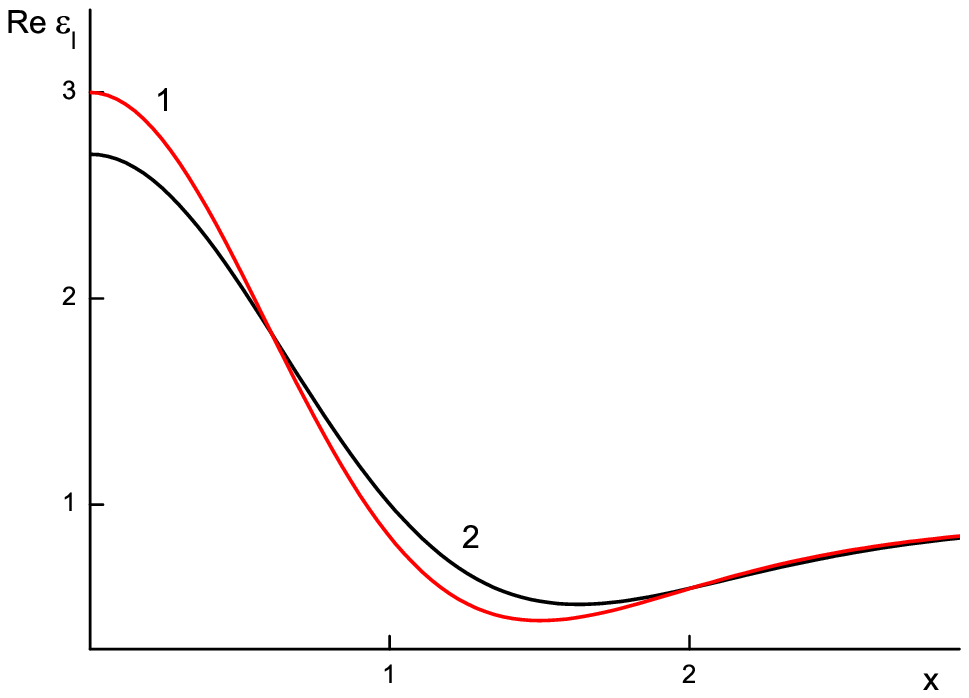}
\hfill
\includegraphics[width=0.49\textwidth]{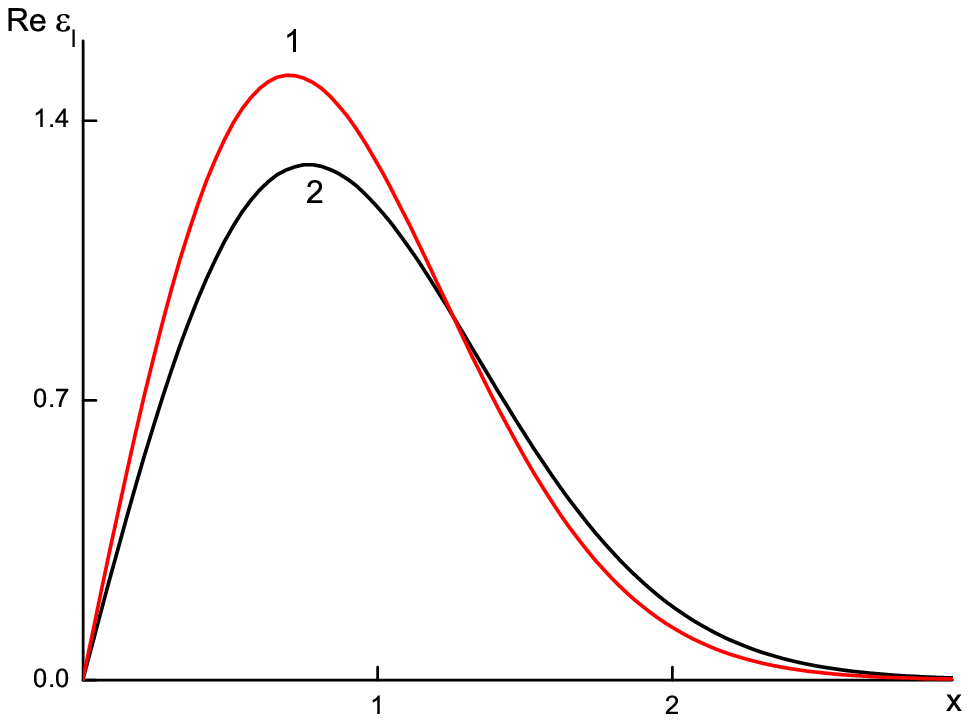}
\\
\parbox[t]{0.47\textwidth}{\hspace{3cm}{Fig. 7.}}
\hfill
\parbox[t]{0.47\textwidth}{\hspace{3cm}{Fig. 8.}}
\end{center}
\end{figure}
\begin{center}
\bf 7. {Conclusion}
\end{center}
In the present work the exact expression for the longitudinal
dielectric permettivity of non--degenerate Maxwellian plasma with
the account of quantum effects is received.
The kinetic Wigner -- Vlasov -- Boltzmann equation
with the collision  integral in the relaxation form
of $\tau$--model in coordinate space is used.

It is shown, that in the limit, when
Planck's constant tends to zero, the received expression
passes into the classical formula of the longitudinal dielectric
permettivity of Maxwellian plasma.
Various special cases of the dielectric permettivity are
investigated. The Lindhard's dielectric function for collisionless
plasma is presented.  It is shown that in the long-wave limit
dielectric permettivity passes into the known Drude formula.
Static limits ($\omega\to 0$)
for dielectric permettivity for collisionless,
and for collision plasmas as well are found.

Plasma oscillations are studied in the long-wave limit and
near to the plasma resonance. The expression for the plasma oscillation
frequence in long-wave limit containing quantum parametre
$Q =\hbar \omega_p/(\varkappa T)$ is received.
Damping of the plasma oscillations is investigated.
The formula for decrement of damping at $\hbar\to 0$ coincide with
classical, and when collision frequency
is equal to zero, the last formula passes into the known Landau damping
formula.

Comparison with classical Mermin's result for dielectric permettivity
is presented.
We will notice, that Mermin's formula is received with use of the
relaxation kinetic equation in space of
momentum. For collisionless plasma the formula deduced in this work,
and Mermin's formula pass into the same Lindhard's formula.

On Figs. 1 and 2 graphic dependence on the quantity $q$ in the case
$x_p=1,\; y=0.1$ for real (fig. 1) and imaginary (fig. 2) parts
of longitudinal dielectric permettivity are presented.
Curves of $1,2,3$ correspond to values of parameter $x=1,\;0.7,\;1.3$.

On Figs. 3 and 4 graphic dependence on the quantity $x$ in the case
$x_p=1,\; y=0.1$ for real (fig. 3) and imaginary (fig. 4) parts
of longitudinal dielectric permettivity are presented.
Curves of $1,2,3$ correspond to values of parameter $q=0.5,\; 0.6,\;0.7$.

On Figs. 5 and 6 graphic dependence on the quantity $x$ in the case
$x_p=10,\; y=0.01, q=1$ for real (fig. 5) and imaginary (fig. 6) parts
of longitudinal dielectric permettivity are presented.
Curves of $1$ correspond to classical plasma, and curves $2$
correspond to quantum plasma.

\begin{figure}[th]
\begin{center}
\includegraphics[width=0.47\textwidth]{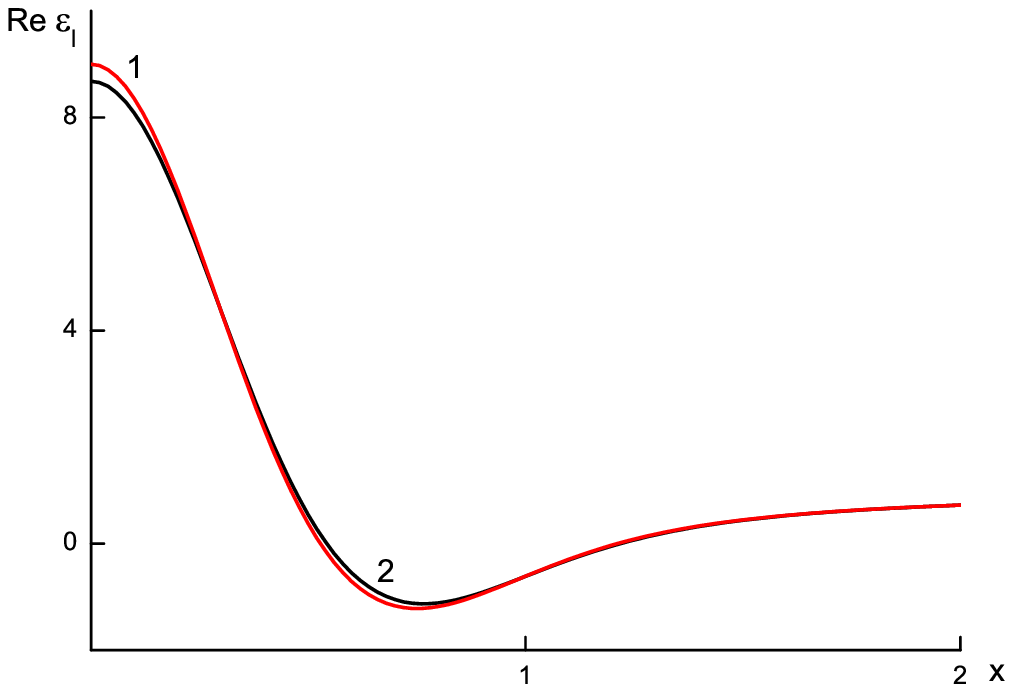}
\hfill
\includegraphics[width=0.47\textwidth]{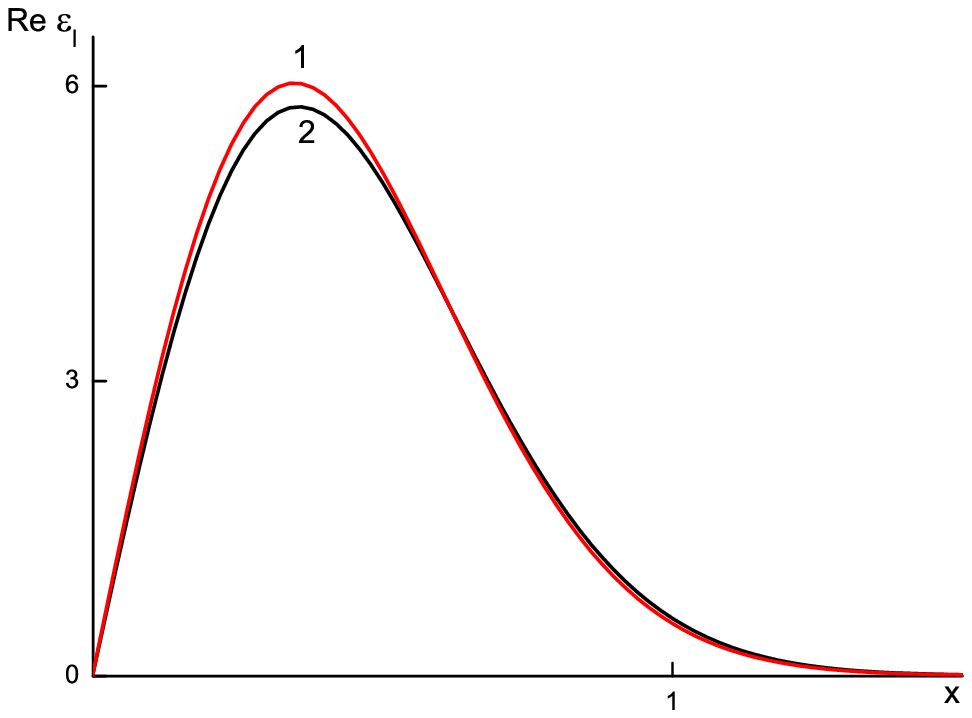}
\\
\parbox[t]{0.47\textwidth}{\hspace{3cm}{Fig. 9.}}
\hfill
\parbox[t]{0.47\textwidth}{\hspace{3cm}{Fig. 10.}}
\end{center}
\end{figure}
\begin{figure}[ht]
\begin{center}
\includegraphics[width=0.47\textwidth]{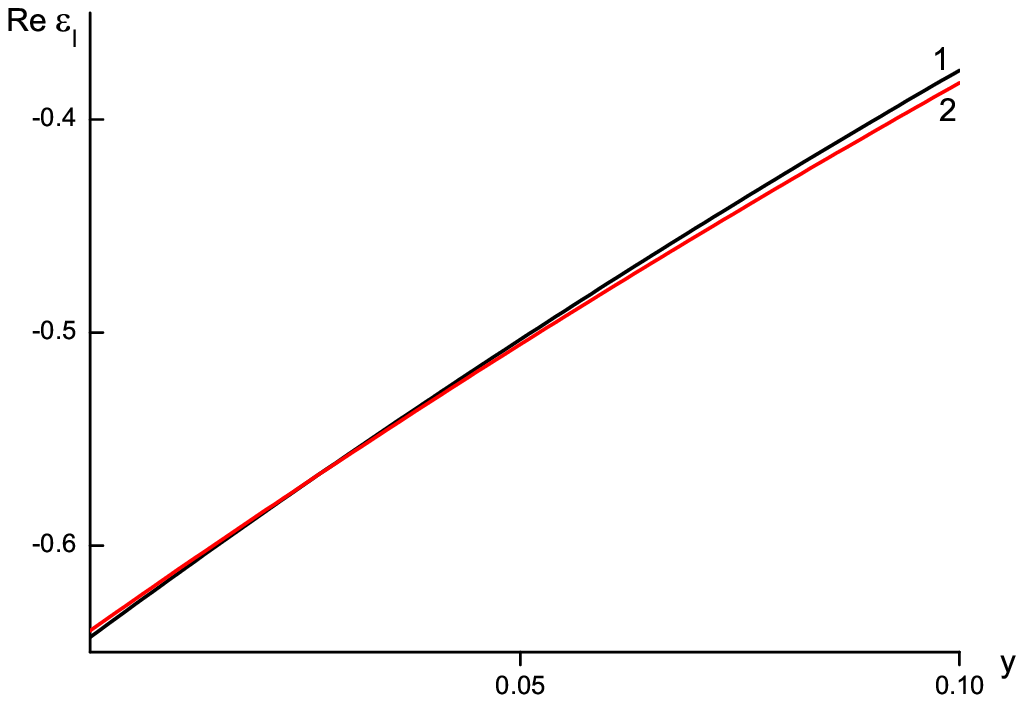}
\hfill
\includegraphics[width=0.47\textwidth]{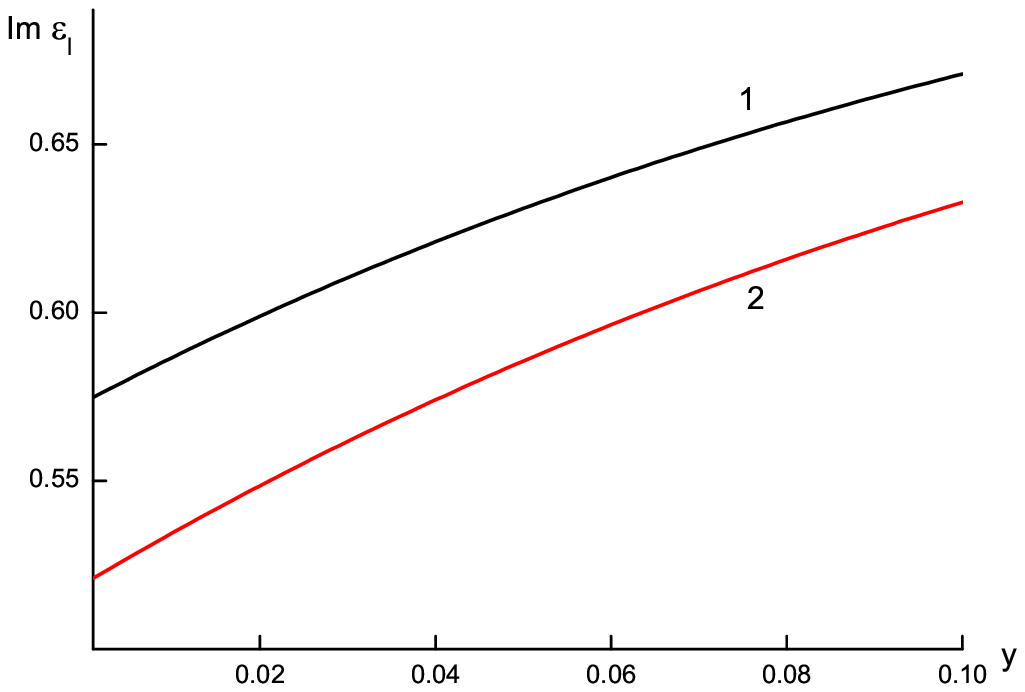}
\\
\parbox[t]{0.47\textwidth}{\hspace{3cm}{Fig. 11.}}
\hfill
\parbox[t]{0.47\textwidth}{\hspace{3cm}{Fig. 12.}}
\end{center}
\end{figure}
\begin{figure}[ht]
\begin{center}
\includegraphics[width=0.47\textwidth]{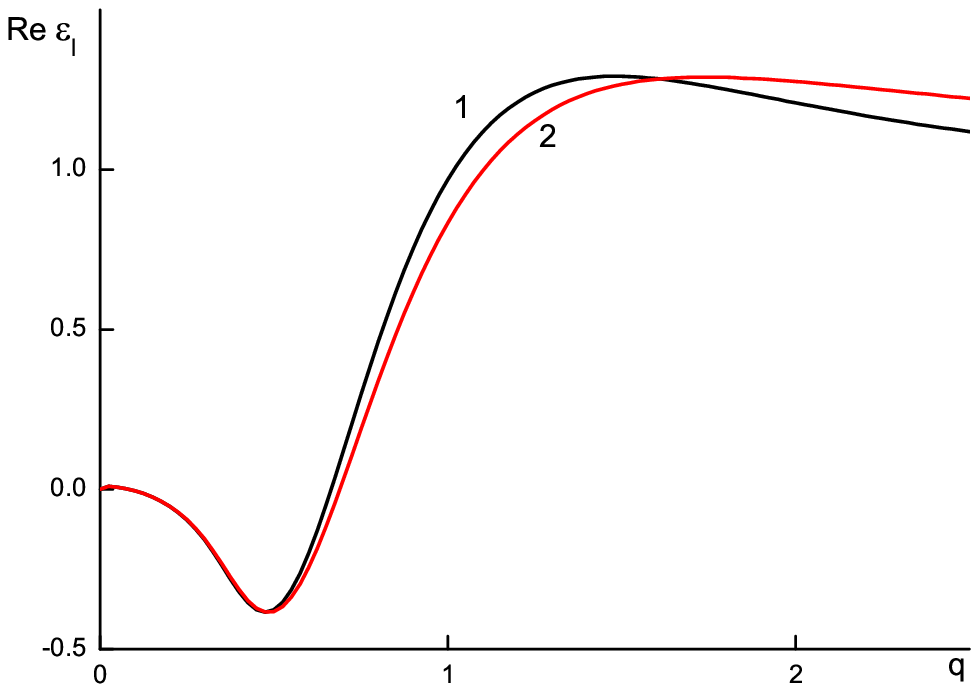}
\hfill
\includegraphics[width=0.47\textwidth]{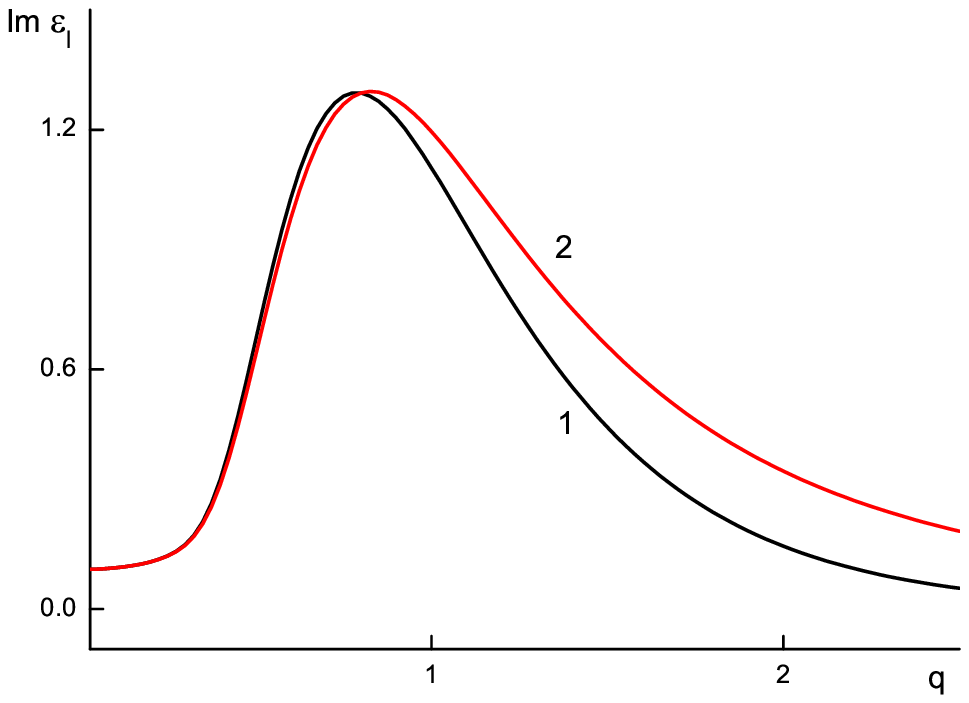}
\\
\parbox[t]{0.47\textwidth}{\hspace{3cm}{Fig. 13.}}
\hfill
\parbox[t]{0.47\textwidth}{\hspace{3cm}{Fig. 14.}}
\end{center}
\end{figure}

On fig. 7 and 8 graphic dependences for real (fig. 7)
and imaginary (fig. 8) parts of longitudinal
dielectric permettivity  on the quntity $x$ for the case
$x_p=1,\; y=0.01,\; q=1$ are presented.
Curves $1$ correspond to quantum plasma, and
curves $2$ correspond to classical plasma.

On fig. 9 and 10 graphic dependences for real (fig. 9)
and imaginary (fig. 10) parts of longitudinal
dielectric permettivity are presented.
These parts depend on quantity $x$
for the case $x_p=1,\; y=0.01,\; q=0.5$.
Curves $1$ correspond to quantum plasma, and
curves $2$ correspond to classical plasma.

On fig. 11 and 12 graphic dependences for real (fig. 11)
and imaginary (fig. 12) parts of longitudinal
dielectric permettivity are presented.
These parts depend on quantity
$y\;(10^{-5}<y<10^{-1})$ for the case $x_p=1,\; x=1,\; q=0.5$.
Curves $1$ correspond to quantum plasma, and
curves $2$ correspond to classical plasma.

On fig. 13 and 14 graphic dependences for real (fig. 13)
and imaginary (fig. 14) parts of longitudinal
dielectric permettivity are presented.
These parts depend on quantity $q\;(0<y<2.5)$
for the case $x_p=1, \; x=1, \; y=0.1$.
Curves $1$ correspond to quantum plasma, and
curves $2$ correspond to classical plasma.

\end{document}